\documentclass[10pt, twocolumn, twoside]{IEEEtran}
\usepackage{amssymb}
\usepackage{amsfonts}

\usepackage{color}

\usepackage{cite}

\ifCLASSINFOpdf
  \usepackage[pdftex]{graphicx}
  \graphicspath{{./Figures/}{./pdf/}{./jpeg/}}
\else
\fi
\usepackage[cmex10]{amsmath}
\usepackage{array}
\usepackage{adjustbox}          
\usepackage{multirow}           
\usepackage{threeparttable}     

\usepackage{stfloats}
\makeatletter
\newcommand{\raisemath}[1]{\mathpalette{\raisem@th{#1}}}
\newcommand{\raisem@th}[3]{\raisebox{#1}{$#2#3$}}
\makeatother

\begin{document}
%
\title{Eigendecomposition-Based Partial FFT Demodulation for Differential OFDM in Underwater Acoustic Communications}
%
%
%

\author{Jing~Han,~\IEEEmembership{Member,~IEEE,}
        Lingling Zhang,~\IEEEmembership{Member,~IEEE,}
        Qunfei~Zhang,~\IEEEmembership{Member,~IEEE}
        and~Geert~Leus,~\IEEEmembership{Fellow,~IEEE}
\thanks{J. Han, L. Zhang and Q. Zhang are with the School of Marine Science and Technology, Northwestern Polytechnical University, Xi'an 710072, China (e-mail: hanj, llzhang, zhangqf@nwpu.edu.cn).}
\thanks{G. Leus is with the Faculty of Electrical Engineering, Mathematics and Computer Science, Delft University of Technology, 2826 CD Delft, The Netherlands (e-mail: g.j.t.leus@tudelft.nl).}}
\maketitle

\begin{abstract}
Differential orthogonal frequency division multiplexing (OFDM) is practically attractive for underwater acoustic communications since it has the potential to obviate channel estimation.
However, similar to coherent OFDM, it may suffer from severe inter-carrier interference over time-varying channels.
To alleviate the induced performance degradation, we adopt the newly-emerging partial FFT demodulation technique in this paper and propose an eigendecomposition-based algorithm to compute the combining weights.
Compared to existing adaptive methods, the new algorithm can avoid error propagation and eliminate the need for parameter tuning. Moreover, it guarantees global optimality under the narrowband Doppler assumption, with the optimal weight vector of partial FFT demodulation achieved by the eigenvector associated with the smallest eigenvalue of the pilot detection error matrix. Finally, the algorithm can also be extended straightforwardly to perform subband-wise computation to counteract wideband Doppler effects.
\end{abstract}




\section{Introduction}
%
%
%
%

\IEEEPARstart{C}{urrently}, underwater acoustic (UWA) communication has been widely chosen as a standard solution for applications of oceanographic data collection. This can be attributed to the relatively low attenuation of acoustic waves in water compared to that of its electromagnetic or optical counterparts \cite{Lanbo&Shengli&Jun-Hong_WCommMCom_2008}.
However, the performance of UWA communication systems may be severely limited by two distortion effects of the UWA channels, namely, multipath spread and Doppler shift.
Specifically, the multipath spread of the UWA channels is usually on the order of tens of milliseconds, which makes channel estimation and equalization very challenging to handle \cite{Stojanovic_IEEEJOE_1996, Stojanovic&Preisig_IEEEMCOM_2009}.
On the other hand, due to the low velocity of acoustic waves (nominally 1500 m/s), the Doppler shift measured by the normalized
carrier frequency offset is often on the order of $10^{-4}$ in UWA channels with mobile transceivers, which is several orders of magnitude greater than that in wireless radio channels \cite{Yahong&Chengshan&TCYang_PHYCOMM_2010, Stojanovic&Preisig_IEEEMCOM_2009}.

To combat the long delay spread efficiently and achieve high-rate transmission over UWA channels, orthogonal frequency division multiplexing (OFDM) is a favorable modulation scheme. This is because OFDM can eliminate inter-symbol interference (ISI) introduced by a frequency-selective channel by transforming it into a set of parallel frequency-flat channels, and therefore enables simple one-tap equalization for each subcarrier \cite{Zhendao&Giannakis_IEEEMSP_2000}.
However, it is well known that the performance of OFDM systems may be significantly degraded over time-varying channels, where the orthogonality among subcarriers is no longer valid \cite{shengli&Zhaohui_OFDMbook_2014}.
To cope with the resulting inter-carrier interference (ICI), existing methods can be divided into two categories, according to their execution order relative to the OFDM demodulation based on the fast Fourier transform (FFT) \cite{Aval&Stojanovic_IEEEJOE_2015}:
\begin{itemize}
  \item Post-FFT methods are performed in the frequency domain (after OFDM demodulation).
  By recognizing that ICI in OFDM is analogous to ISI in single-carrier modulation (SCM), block equalization \cite{Rugini&Banelli&Leus_IEEEJCOML_2005, KunFang&Rugini&Geert_IEEEJSP_2008} and serial equalization \cite{Schniter_IEEEJSP_2004, Tu&Fertonani&Duman&Stojanovic_IEEEJOE_2011, Jianzhong&Shengli&Jie_IEEEJSTSP_2011} have been utilized to mitigate its effects.

  \item Pre-FFT methods are performed in the time domain (before OFDM demodulation).
  Among them, simple carrier frequency offset (CFO) compensation has been adopted in \cite{Baosheng&Shengli&Stojanovic_IEEEJOE_2008, Kang&Iltis_IEEEJJSAC_2008}. Besides, a newly-emerging technique referred to as partial FFT demodulation was also proposed in \cite{Yerramalli&Stojanovic&Mitra_SPAWC_2010, Yerramalli&Stojanovic&Mitra_IEEEJSP_2012}, which can counteract more complicated time variations within each OFDM block.

\end{itemize}

\IEEEpubidadjcol

More specifically, the partial FFT demodulation algorithm in \cite{Yerramalli&Stojanovic&Mitra_IEEEJSP_2012} focuses on coherent OFDM detection.
It divides each OFDM block into several non-overlapping subblocks, and then performs a weighted combining of the corresponding partial FFT outputs.
Mathematically, it is equivalent to imposing a step-wise window in the time domain for each subcarrier before OFDM demodulation, and hence has the capability to alleviate ICI \cite{Yerramalli&Stojanovic&Mitra_IEEEJSP_2012}. This idea has also been extended to multiple-input multiple-output OFDM systems in \cite{Jing&Lingling&Geert_IEEESPL_2016}.
However, these partial FFT demodulation algorithms have to be coupled with channel estimation in the frequency domain. As a result, given the complex time-varying nature of the UWA channels, they may suffer a severe performance loss due to the channel estimation error.

Alternatively, differential OFDM is an attractive scheme since it has the potential to eliminate the need for channel estimation. However, similar to its coherent counterpart, differential OFDM is susceptible to channel time variations, and ICI mitigation has to be performed.
To this end, the post-FFT methods are not favorable, because they necessitate channel estimation for ICI equalization which contradicts the aim of introducing differential coding.
On the other hand, it can be much easier in this scenario to exempt the pre-FFT methods from explicit channel estimation.

The emphasis of this paper is on the partial FFT demodulation for differential OFDM systems.
So far, there has been not much research on this issue; and to the best of our knowledge, \cite{Stojanovic_SAM_2010} is the first attempt to investigate its feasibility.
As an important variant of its original version for coherent OFDM in~\cite{Yerramalli&Stojanovic&Mitra_IEEEJSP_2012}, a stochastic gradient algorithm was designed in~\cite{Stojanovic_SAM_2010} to minimize the differential mean squared error (MSE) and update the partial FFT weights across subcarriers recursively. The follow-up work in \cite{Aval&Stojanovic_IEEEJOE_2015} further enhanced the adaptive algorithm and used it for a multichannel receiver to exploit the spatial diversity gain.
These previous works have shown the validity of partial FFT demodulation in compensating for the channel time variation in differential OFDM systems. However, since the differential MSE is not a convex function of the partial FFT weights, there will be no guarantee for these adaptive algorithms to converge close to the global optimum. Moreover, their performances will suffer from the error propagation effect caused by deep fading at some subcarriers, and from the sensitivity to the choice of parameters, such as the step size.

To solve the above problems, an eigendecomposition-based partial FFT demodulation algorithm is proposed for differential OFDM in this paper, which has the following features:
\begin{itemize}
  \item The algorithm computes the combining weights of partial FFT demodulation in a non-adaptive manner based on pilot symbols. As such, it can avoid error propagation and eliminate the need for parameter tuning.
  \item When the Doppler effect at the receiver (after front-end resampling) can be approximately modeled as narrowband, the algorithm guarantees global optimality. It is shown that the optimal weight vector of partial FFT demodulation is the eigenvector associated with the smallest eigenvalue of the pilot detection error matrix.
  \item The algorithm can also be extended straightforwardly to the case where wideband Doppler effects cannot be ignored. In this case, subband-specific weight vectors are produced at the expense of an increased pilot overhead.
\end{itemize}

The remainder of this paper is organized as follows. In Section II, we present the differential OFDM signal model and the UWA channel model. In Section III, we describe the proposed partial FFT demodulation algorithm in detail, based on which the numerical simulation results are then presented in Section IV. Finally, conclusions are drawn in Section V.

\textit{Notation}: ${\left(  \cdot  \right)^*}$ stands for conjugate, ${\left(  \cdot  \right)^T}$ for transpose, ${\left(  \cdot  \right)^H}$ for Hermitian transpose.
We reserve $\left|  \cdot  \right|$ for the absolute value, $\left\|  \cdot  \right\|$ for the Euclidean norm and $\otimes$ for the Kronecker product.
Also, we use ${\bf{0}}_M$, ${\bf{1}}_M$, ${\bf{I}}_M$ and ${{\bf{i}}_M}\left( m \right)$ to represent the $M \times 1$ all-zero vector, the $M \times 1$ all-one vector, the $M \times M$ identity matrix and the $m$th column of ${\bf{I}}_M$, respectively.
In addition, ${\bf{F}}_K$ denotes the $K \times K$ unitary DFT matrix, and ${\rm{diag}}\left\{ {\bf{x}} \right\}$ denotes a diagonal matrix with $\bf{x}$ on its diagonal.

\section{System Model}

\begin{figure*}[!t]
\centering
\includegraphics[width=7in]{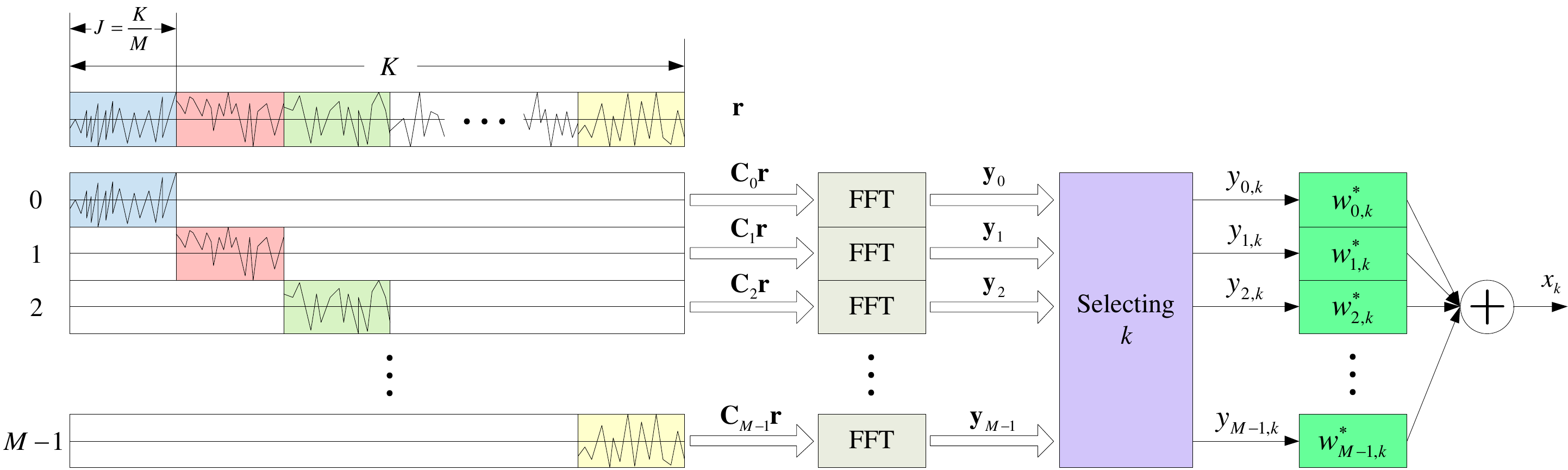}
\caption{The processing of partial FFT demodulation.}
\label{Figure_1}
\end{figure*}

Consider a differential OFDM system with $K$ subcarriers.
A simple scalar differential encoding scheme is applied here as in \cite{Aval&Stojanovic_IEEEJOE_2015, Stojanovic_SAM_2010}.
To be specific, let $b_k$ and $d_k$ denote the original information symbol and the differentially coded symbol modulated on the $k$th subcarrier, respectively. Assuming that both $b_k$ and $d_k$ are drawn from the same normalized $Q$-ary phase-shift keying (PSK) constellation set $\mathcal{A} = \left\{ {{a_0}, {a_1}, \ldots ,{a_{Q-1}}} \right\}$ with ${a_q} = {e^{jq/Q}}$, $q=0, 1, \ldots, Q-1$, the generation of the differentially coded symbol $d_k$ follows the recursion
\begin{IEEEeqnarray}{c}
    d_k = \left\{ \,
    \begin{IEEEeqnarraybox}[][c]{l?l}
        \IEEEstrut
        {b_k d_{k-1},}      &       {1 \le k \le K-1}, \\
        {a_0,}              &       {k=0},
        \IEEEstrut
    \end{IEEEeqnarraybox}
\right.
\label{eqn_diff_encoding}
\end{IEEEeqnarray}
where $d_0$ is the known initial symbol in order to start the encoding process.
Furthermore, by collecting all $K$ differentially coded symbols and defining ${\bf{d}} = [d_0, d_1, \ldots, d_{K-1}]^T$, the baseband differential OFDM block ${\bf{s}} = {[{s_0},{s_1}, \ldots ,{s_{K - 1}}]^T}$ can be expressed as
\begin{IEEEeqnarray}{rCl}
{\bf{s}} = {\bf{F}}_K^H {\bf{d}}.
\label{eqn_s}
\end{IEEEeqnarray}
Finally, after insertion of a cyclic prefix (CP), the differential OFDM block is pulse shaped to the corresponding continuous-time signal, upconverted to the carrier frequency, and then transmitted through the UWA channel.

At the receiver, since UWA communication signals are wideband in nature, front-end resampling is often required to mitigate the time compression/dilation induced by time-varying channels \cite{Sharif&Neasham&Hinton_IEEEJOE_2000}. After this operation, the residual Doppler effect can usually be treated as narrowband.
For instance, under the channel assumption that path amplitudes are constant and a common Doppler scale is shared among all paths, it has been shown that the effect of time variation in wideband signals approximately reduces to a CFO after resampling \cite{Baosheng&Shengli&Stojanovic_IEEEJOE_2008}.
Based on that, in this paper, we adopt a more general model in \cite{Stojanovic&Catipovic&Proakis_IEEEJOE_1994, Yahong&Chengshan&TCYang_PHYCOMM_2010, JHan&Leus_IEEEJOE_EA}, which further eliminates the single-frequency restriction and uses a narrowband phase distortion to represent the post-resampling Doppler effect. Correspondingly, the received differential OFDM block (after downconversion and CP removal) can be written as
\begin{IEEEeqnarray}{rCl}
{\bf{r}} = {\bf{G}} {\bf{\tilde H}} {\bf{s}} + {\bf{z}},
\label{eqn_r}
\end{IEEEeqnarray}
where $\bf{\tilde H}$ is the $K \times K$ circulant channel matrix with first column equal to the channel impulse response (CIR) vector ${\bf{h}} = {\left[ {{h_0},{h_1}, \ldots {h_L}} \right]^T}$ appended by $K - L - 1$ zeros; ${\bf{G }} = {\rm{diag}} \{ {\bf{g}} \}$ is the time-varying phase matrix with ${\bf{g}} = [ {{e^{j{\theta _0}}},{e^{j{\theta _1}}}, \ldots ,{e^{j{\theta _{K - 1}}}}} ]^T$ on its diagonal;
${\bf{z}}$ is the noise vector with independent and identically distributed (i.i.d.) entries of zero mean and variance ${\sigma}^2$.

For clarity, let us first consider the time-invariant channel case where ${\bf{G}} = {\bf{I}}_K$.
Since conventional OFDM demodulation uses a single $K$-point FFT, we define ${\bf{x}} = {\bf{F}}_K {\bf{r}}$ and ${\bf{u}} = {\bf{F}}_K {\bf{z}}$ accordingly. Recalling the fact that circulant channel matrices can be diagonalized by FFT matrices, we obtain
\begin{IEEEeqnarray}{rCl}
{\bf{x}} = {\bf{H}} {\bf{d}} + {\bf{u}},
\label{eqn_x}
\end{IEEEeqnarray}
where ${\bf{H}} = {\rm{diag}}\left\{ {\left[ {{H_0},{H_1}, \ldots ,{H_{K - 1}}} \right]} \right\}$ with diagonal entries ${H_k} = \sum\nolimits_{l = 0}^L {{h_l}{e^{ - j\left( {2\pi /K} \right)lk}}}$ for $k = 0, \ldots, K-1$.
Then, under the assumption that the channel does not change much over two consecutive subcarriers, i.e., ${H_{k-1}} \approx {H_k}$, it yields
\begin{IEEEeqnarray}{rCl}
{x_k} = {x_{k-1}} {b_k} + {v_k},
\label{eqn_xk_diff}
\end{IEEEeqnarray}
for $1 \le k \le K-1$, where ${v_k} = {u_k} - {u_{k - 1}}{b_k}$ is the differential noise term.
It is easy to verify that the variance of ${v_k}$ is $2 {\sigma}^2$, twice that of the original noise samples in the vector $\bf{z}$, which
corresponds to the well-known 3-dB performance loss of differential detectors, relative to coherent detectors.

Based on (\ref{eqn_xk_diff}), the maximum-likelihood (ML) detector for ${b_k}$ can be represented by
\begin{IEEEeqnarray}{rCl}
{{\check b}_k} = \arg \mathop {\min }\limits_{{b} \in {\mathcal{A}}} \left| {{x_k} - {x_{k - 1}}{b}} \right|^2.
\label{eqn_det1_ML}
\end{IEEEeqnarray}
Since the cardinality of $\mathcal{A}$ is $Q$, the complexity of optimal ML decoding for a differential OFDM block is about $\mathcal{O}(QK)$.
Moreover, from (\ref{eqn_xk_diff}), we can readily arrive at another method to detect the information symbol ${b_k}$, i.e.,
\begin{IEEEeqnarray}{rCl}
{{\hat b}_k} = \frac{{{x_k}}}{{{x_{k - 1}}}},
\qquad
{{\check b}_k} = {\rm{dec}} \{ {{\hat b}_k} \},
\label{eqn_det2_Stojanovic}
\end{IEEEeqnarray}
where ${\rm{dec}} \{ \cdot \}$ maps the point to the nearest constellation symbol.
The complexity of this latter method is also linear in the number of subcarriers.

However, over time-varying UWA channels, the differential OFDM detection is much more complicated.
This is a result of the presence of the matrix $\bf{G}$ in (\ref{eqn_r}), and correspondingly the diagonal channel matrix $\bf{H}$ in (\ref{eqn_x}) needs to be pre-multiplied by a circulant matrix ${\bf{\tilde G}} = {{\bf{F}}_K} {\bf{G}} {{\bf{F}}_K^H}$.
In this case, the demodulated symbols in $\bf{x}$ are no longer decoupled, and ICI elimination has to be performed before the differential decoding with (\ref{eqn_det1_ML}) or (\ref{eqn_det2_Stojanovic}). To cope with it, we focus here on the partial FFT demodulation for differential OFDM, which will be discussed in detail in the following section.

\section{Partial FFT Demodulation}

\subsection{The Existing Algorithms}
The processing of partial FFT demodulation is depicted in Fig.~1. Specifically, the OFDM block is divided into $M$ subblocks in the time domain by applying $M$ non-overlapping rectangular windows, and the $m$th window is defined by
\begin{IEEEeqnarray}{rCl}
{{\bf{c}}_m} = {{\bf{i}}_M}\left( m \right) \otimes {{\bf{1}}_{J}},
\label{eqn_cm}
\end{IEEEeqnarray}
where $m = 0, \ldots, M-1$ and $J = K/M$.
Then, unlike the conventional OFDM demodulation which performs a single FFT on the whole OFDM block, the partial FFT demodulation operates on each subblock in parallel. Accordingly, the demodulation output of the $m$th subblock takes the form
\begin{IEEEeqnarray}{rCl}
{{\bf{y}}_m} = {{\bf{F}}_K}{{\bf{C}}_m}{\bf{r}},
\label{eqn_ym}
\end{IEEEeqnarray}
with ${{\bf{C}}_m} = {\rm{diag}}\left\{ {{{\bf{c}}_m}} \right\}$.
Moreover, in the frequency domain, weighted combining is performed on each subcarrier to mitigate ICI. Let us define ${{{\bf{w}}}_k} = {\left[ {{w_{0,k}},{w_{1,k}}, \ldots ,{w_{M - 1,k}}} \right]^T}$ as the partial FFT weight vector and ${{{\bf{\bar y}}}_k} = {\left[ {{y_{0,k}},{y_{1,k}}, \ldots ,{y_{M - 1,k}}} \right]^T}$ where $y_{m,k}$ is the $k$th entry of the vector ${{\bf{y}}_m}$. Then, the final demodulated symbol on the $k$th subcarrier can be written as
\begin{IEEEeqnarray}{rCl}
{x_k} = {\bf{w}}_k^H{{{\bf{\bar y}}}_k}.
\label{eqn_xk_combining}
\end{IEEEeqnarray}

As for computing the weight vector ${{{\bf{w}}}_k}$, the differential OFDM systems in \cite{Aval&Stojanovic_IEEEJOE_2015, Stojanovic_SAM_2010} are based on the detector in (\ref{eqn_det2_Stojanovic}), and adaptive algorithms are designed to minimize the differential MSE, i.e.,
\begin{IEEEeqnarray}{rCl}
E \left\{ {{{\left| {{{\xi}_k}} \right|}^2}} \right\}
& = & E \left\{ {{{\left| {{b_k} - {{\hat b}_k}} \right|}^2}} \right\}      \nonumber   \\
& \approx & E\left\{ {{{\left| {{b_k} - \frac{{{\bf{w}}_k^H{{{\bf{\bar y}}}_k}}}{{{\bf{w}}_{k}^H{{{\bf{\bar y}}}_{k - 1}}}}} \right|}^2}} \right\}.
\label{eqn_MMSE}
\end{IEEEeqnarray}
In the second equation, it has been assumed that the channel time variation is highly correlated over adjacent subcarriers, and so is the weight vector, i.e., ${{\bf{w}}_k} \approx {{\bf{w}}_{k - 1}}$.

Based on (\ref{eqn_MMSE}), the original stochastic gradient algorithm in \cite{Stojanovic_SAM_2010} updates the weight vector recursively across $K$ subcarriers. When it operates in decision-directed mode, $b_k$ in (\ref{eqn_MMSE}) is actually replaced by ${\check b}_k$ in (\ref{eqn_det2_Stojanovic}). The algorithm may thus suffer from error propagation due to the deep fading in the channel frequency response.
To this end, an improved stochastic gradient algorithm was presented in \cite{Aval&Stojanovic_IEEEJOE_2015}, which uses a scaled gradient combined with a thresholding method in order to mitigate abrupt weight changes caused by decision errors.
However, unlike the classical MSE in coherent OFDM which is a convex quadratic function of the weight vector, the differential MSE in (\ref{eqn_MMSE}) is a non-convex function of ${{{\bf{w}}}_k}$. Therefore, these algorithms cannot guarantee global optimality.
In addition, the performance of these adaptive algorithms depends heavily on the choice of parameters, such as the step size, which may preclude them from practical use.

\subsection{The Eigendecomposition-Based Algorithm}

To solve the above problems, an alternative algorithm is proposed for computing the partial FFT weights based on eigendecomposition.
Instead of using the detector in (\ref{eqn_det2_Stojanovic}) as \cite{Aval&Stojanovic_IEEEJOE_2015, Stojanovic_SAM_2010}, we here employ the ML detector in (\ref{eqn_det1_ML}).
Moreover, for ease of presentation, we assume at this point that the weight vector remains constant over all $K$ subcarriers, i.e., ${\bf{w}} = {{\bf{w}}_0} =  \cdots  = {{\bf{w}}_{K - 1}}$, which is justified by the narrowband Doppler effect after resampling in (\ref{eqn_r}).
Then, by inserting (\ref{eqn_xk_combining}) into (\ref{eqn_det1_ML}), the ML detector can be reformulated as
\begin{IEEEeqnarray}{rCl}
{{\check b}_k}
& = & \arg \mathop {\min }\limits_{b \in {\mathcal{A}}} {\left| {{{\bf{w}}^H}{{{\bf{\bar y}}}_k} - {{\bf{w}}^H}{{{\bf{\bar y}}}_{k - 1}}b} \right|^2}  \nonumber   \\
& = & \arg \mathop {\min }\limits_{b \in {\mathcal{A}}} {{\bf{w}}^H}{{\bf{R}}_k} (b) {\bf{w}},
\label{eqn_bk_wRw}
\end{IEEEeqnarray}
where we have defined the rank-$1$ error matrix
\begin{IEEEeqnarray}{rCl}
{{\bf{R}}_k}\left( b \right)
& = & \left( {{{\bf{\bar y}}}_k} - {{{\bf{\bar y}}}_{k - 1}}b \right) \left( {{{\bf{\bar y}}}_k} - {{{\bf{\bar y}}}_{k - 1}}b \right)^H     \nonumber \\
& = & {{\bf{e}}_k}\left( b \right){\bf{e}}_k^H\left( b \right).
\label{eqn_Rk}
\end{IEEEeqnarray}

The proposed algorithm uses $I$ pilot symbols for computing the combining weights of partial FFT demodulation in a non-adaptive manner, by which the requirement of parameter tuning can be eliminated. Additionally, it is observed in (\ref{eqn_bk_wRw}) that, except for the pathological case when ${\bf{w}} = {{\bf{0}}_{M}}$, the ML symbol decision does not depend on the norm of the weight vector. We can thus simplify the weight computation by fixing $\left\| {\bf{w}} \right\|$ to any nonzero value.
To be specific, define the index set of pilot symbols ${{\mathcal{K}}_{\rm{P}}} = \left\{ {{k_0},{k_1}, \ldots ,{k_{I - 1}}} \right\}$, i.e., $\left\{ {{b_k} \, | \, k \in {{\mathcal{K}}_{\rm{P}}}} \right\}$ are known to the receiver, and thus also the corresponding $M \times I$ matrix ${{\bf{E}}_{\rm{P}}} = \left[ {{{\bf{e}}_{{k_0}}}({b_{{k_0}}}),{{\bf{e}}_{{k_1}}}({b_{{k_1}}}), \ldots ,{{\bf{e}}_{{k_{I - 1}}}}({b_{{k_{I - 1}}}})} \right]$. The partial FFT weight vector can then be obtained by minimizing the total error energy of differential detection on all pilot symbols, i.e., by solving the following optimization problem
\begin{IEEEeqnarray}{lCl}
{\mathop {\min }\limits_{\bf{w}}}       &   \quad   &     {{\bf{w}}^H}{{\bf{R}}_{\rm{P}}}{\bf{w}}       \label{eqn_w_opt_prob}    \\
{\rm{s.t.}}                             &   \quad   &     {\left\| {\bf{w}} \right\| = \sqrt M},        \nonumber
\end{IEEEeqnarray}
where ${{\bf{R}}_{\rm{P}}}$ is the pilot detection error matrix defined as
\begin{IEEEeqnarray}{rCl}
{{\bf{R}}_{\rm{P}}}
= \sum\nolimits_{k \in {{\mathcal{K}}_{\rm{P}}}} {{{\bf{R}}_k}\left( {{b_k}} \right)}
= {{\bf{E}}_{\rm{P}}}{\bf{E}}_{\rm{P}}^H.
\label{eqn_RP}
\end{IEEEeqnarray}

As can be observed, the optimization problem in (\ref{eqn_w_opt_prob}) is not convex either. However, since ${\bf{R}}_{\rm{P}}$ is a Hermitian matrix, global optimality can actually be achieved in this special case~\cite{Matrix_Meyer_2000_inbook_MinMaxEigen}. The minimum value of (\ref{eqn_w_opt_prob}) is $M {\lambda}_{\min }$, where ${\lambda}_{\min }$ is the smallest eigenvalue of ${\bf{R}}_{\rm{P}}$. And the optimal partial FFT weight vector can be computed as
\begin{IEEEeqnarray}{rCl}
{\bf{\hat w}}_{\rm{opt}} = \sqrt M {{\bf{v}}_{\min }},
\label{eqn_w_opt_est}
\end{IEEEeqnarray}
with ${{\bf{v}}_{\min }}$ the normalized eigenvector associated with ${\lambda}_{\min }$.
More discussions on this eigendecomposition-based algorithm are presented next.

\subsubsection{Uniqueness of the Optimal Weight Vector}

It can be seen that the conventional single-FFT demodulation corresponds to the feasible point ${\bf{w}} = {{\bf{1}}_{M}}$ in (\ref{eqn_w_opt_prob}).
When the channel is frequency-flat over adjacent subcarriers with time variations and noise absent, it is also an eigenvector associated with the smallest eigenvalue ${\lambda}_{\min } = 0$. This is because ${\bf{R}}_{\rm{P}}{{\bf{1}}_{M}} = {{\bf{0}}_{M}}$, and zero error energy is achieved perfectly here.
We can thus say that the conventional single-FFT demodulation is sufficient to provide the ML performance in this simple case.
In contrasts, over time-varying UWA channels, the all-one vector is probably not optimal and the partial FFT weights have to be computed as (\ref{eqn_w_opt_est}) to mitigate the Doppler-induced ICI. Generally, the total error energy ${{\bf{w}}^H}{{\bf{R}}_{\rm{P}}}{\bf{w}}$ in (\ref{eqn_w_opt_prob}) can no longer be reduced to zero in practical cases.
However, two issues on the uniqueness of ${\bf{\hat w}}_{\rm{opt}}$ need to be noted here:
\begin{itemize}
  \item First, at least $I \ge M$ pilot symbols are required to make ${\bf{E}}_{\rm{P}}$ have full row rank, and thus ${\bf{R}}_{\rm{P}}$ positive definite. Otherwise, ${\bf{R}}_{\rm{P}}$ is rank-deficient, i.e., ${\lambda}_{\rm{min}} = 0$, and all the vectors in its null space may produce zero error energy ${{\bf{w}}^H}{{\bf{R}}_{\rm{P}}}{\bf{w}}$ in (\ref{eqn_w_opt_prob}). No valid weight vector can be determined in this case.

  \item Second, although rarely occurring, theoretically it is also possible that ${\lambda}_{\rm{min}}$ is positive, however, with multiplicity larger than one. In this case, there are multiple orthonormal eigenvectors associated with ${\lambda}_{\rm{min}}$, and the optimal weight vector ${\bf{\hat w}}_{\rm{opt}}$ is not unique. The uncertainty can be easily overcome by reassigning $M$ with a slightly different integer value (provided $I \ge M$ still holds).
\end{itemize}

\subsubsection{Bandwidth Efficiency and Complexity}
Recall that, in coherent OFDM systems, pilot symbols are typically used to estimate the CIR, and thus their number directly depends on the channel delay spread. For instance, in \cite{Baosheng&Shengli&Stojanovic_IEEEJOE_2008}, at least $L+1$ pilot symbols are required in each OFDM block to facilitate block-by-block channel estimation. So, over UWA channels with extended multipath, this may lead to a significant loss of bandwidth efficiency.
In comparison, here for the differential OFDM system, pilot symbols are introduced to compute the partial FFT weights with the requirement that $I \ge M$. Since, in practice, the residual maximum Doppler shift after front-end resampling can usually be confined to less than the subcarrier spacing, typically selecting the number of partial FFT subblocks as $M = 4 \sim 32$ is enough for good Doppler compensation. Therefore,
the pilot overhead incurred will be much smaller than that in coherent OFDM systems.

As for computational complexity, constructing ${\bf{R}}_{\rm{P}}$ in (\ref{eqn_RP}) requires $\mathcal{O} ( M^2I )$ floating-point operations (flops). Furthermore, the eigendecomposition complexity of ${\bf{R}}_{\rm{P}}$ is of order $\mathcal{O} ( M^3 )$ flops.
Although cubic in $M$, as stated previously that we can choose a small value of $M$, and thus $M \ll K$ in practice, the complexity of the proposed algorithm is actually not much and tractable.

\subsubsection{Comparison with Existing Methods}
At first sight, the resulting weight vector of the proposed algorithm seems to be a step-wise estimate of the channel time variation, i.e., ${\bf{\hat g }} = {{\bf{\hat w}}_{\rm{opt}}} \otimes {{\bf{1}}_{J}}$.
It may be reminiscent of the phase correction algorithm designed for SCM systems in \cite{Yahong&Chengshan&TCYang_PHYCOMM_2010}, which
estimates the time average of the phase distortion at each subblock in a decision-directed way.
However, it should be noted that ${\bf{\hat w}}_{\rm{opt}}$ obtained here does not strictly enforce unit-magnitude entries as the definition of ${\bf{g}}$ in~(\ref{eqn_r}). By imposing the relaxed constraint ${\left\| {\bf{w}} \right\| = \sqrt M}$ instead of ${\left| {{w_m}} \right| = 1}$, $m = 0, \ldots, M-1$, where $w_m$ is the $m$th entry of $\bf{w}$, the optimization problem in (\ref{eqn_w_opt_prob}) can be solved much more efficiently. In addition, better performance of ICI mitigation is enabled with the enhanced capability to accommodate other effects of channel time variation that cannot be aggregated into simple phase distortion.

Furthermore, compared with the stochastic gradient algorithms in \cite{Aval&Stojanovic_IEEEJOE_2015, Stojanovic_SAM_2010},
this algorithm can guarantee the finding of a global minimum via eigendecomposition performed in a non-adaptive way.
Therefore, it is free from the problems of premature convergence, error propagation and high sensitivity to parameter choice.

\subsubsection{Extension for Wideband Doppler Effects}

So far, we have assumed narrowband Doppler effects after front-end resampling at the receiver. As shown in (\ref{eqn_r}), the diagonal matrix $\bf{G}$ is used to model the common phase distortion on all $K$ subcarriers.
Therefore, unlike the algorithms in \cite{Aval&Stojanovic_IEEEJOE_2015, Stojanovic_SAM_2010}, which update the weight vector subcarrier-wise,
here the optimal weight vector is computed only once for each OFDM block.
Although this narrowband Doppler model is usually sufficient for practical use, it is worth mentioning that the proposed eigendecomposition algorithm can also be extended straightforwardly to wideband Doppler scenarios, where the discrepancy of the post-resampling Doppler effect across subcarriers cannot be ignored.
Specifically, in this case we divide the bandwidth of the differential OFDM system into $N$ subbands. Each subband spans ${\bar K} = K/N$ subcarriers, among which ${\bar I} \ge M$ pilot symbols are placed. Then, assuming the narrowband Doppler model holds approximately on the subband level, we can again resort to (\ref{eqn_w_opt_est}) to compute the optimal partial FFT weight vectors specific to each subband, by which better performance of Doppler compensation can be expected at the expense of an increase in overhead (since at least $MN$ pilot symbols are needed).

\section{Numerical Simulations}

In this section, numerical simulation results are provided to illustrate the bit error rate (BER) performance of the eigendecomposition-based  partial FFT demodulation. In all following simulations, we investigate a differential OFDM system with $K = 1024$ subcarriers and constellation size $Q = 4$. A total bandwidth of $B = 4096$~Hz at a center frequency of $6$~kHz is used. The subcarrier spacing is $\Delta f = B/K = 4$~Hz and the OFDM block duration is $T = 1/\Delta f = 0.25$~s.
In addition, the UWA channel is assumed to have $L+1 = 48$ taps with uniform power-delay profile, which corresponds to a maximum delay spread of ${\tau}_{\rm{max}} = LT/K \approx 11.5$~ms,
and thus the coherence bandwidth can be coarsely calculated as $B_{\rm{c}} = 1/{\tau}_{\rm{max}} \approx 87.1$~Hz. Since $\Delta f \ll B_{\rm{c}}$, the channel frequency response can be considered constant over neighboring subcarriers, which justifies the differential detection in~(\ref{eqn_det1_ML}) or (\ref{eqn_det2_Stojanovic}).
Furthermore, the channel time variation here is simulated by a post-resampling Doppler scaling factor $a$, and the nonuniform Doppler shift among subcarriers is taken into account explicitly. To be specific, the Doppler shift at the $k$th subcarrier of the OFDM signal is $af_k$, where $f_k$ is the subcarrier frequency.

\begin{figure}[!t]
\centering
\includegraphics[width=3in]{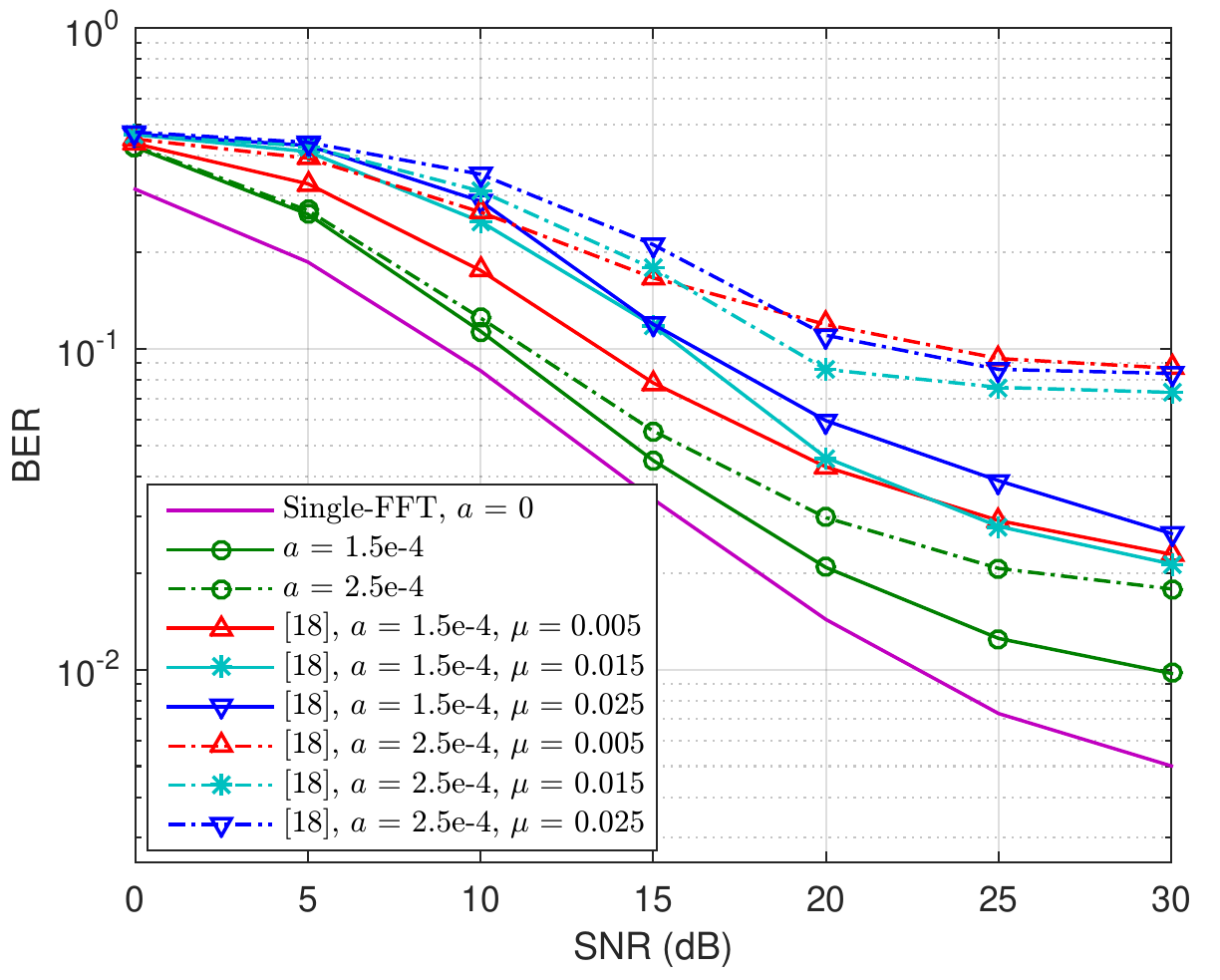}
\vskip -0.7em
\caption{BER performance comparison between the proposed algorithm and the partial FFT demodulation algorithm in \cite{Stojanovic_SAM_2010}.}
\label{fig2}
\vskip -1em
\end{figure}

Fig.~2 compares the BER performance of the proposed eigendecomposition-based algorithm and that of the algorithm in~\cite{Stojanovic_SAM_2010}. Two sets of results, corresponding to Doppler parameter values $a = 1.5 \times 10^{-4}$ and $2.5 \times 10^{-4}$, are presented.
For fairness, both partial FFT demodulation algorithms fix the number of subblocks to $M = 8$ and use $I = 32$ pilot symbols. The only difference is that these pilots are placed with equal spacing $P = K/I$ for the proposed algorithm, i.e., ${\mathcal{K}}_{\rm{P}} = \{ 1, P+1, \ldots, (I-1)P+1\}$, to capture Doppler distortion over the entire frequency band, while continuously assigned at the low frequency end for the algorithm in~\cite{Stojanovic_SAM_2010}, i.e., ${\mathcal{K}}_{\rm{P}} = \{ 1, 2, \ldots, I\}$, to perform initial training.
Moreover, the BER curve of the conventional single-FFT demodulation (i.e., $M=1$) over the time-invariant channel is included as a benchmark.
Compared to that, it can be seen that the performance of the algorithm in \cite{Stojanovic_SAM_2010} suffers a significant degradation, and it is sensitive to the step size $\mu$. On the other hand, the proposed algorithm always produces lower BERs than its adaptive counterpart due to the ability to achieve global optimality and immunity from error propagation.

\begin{figure}[!t]
\centering
\includegraphics[width=3in]{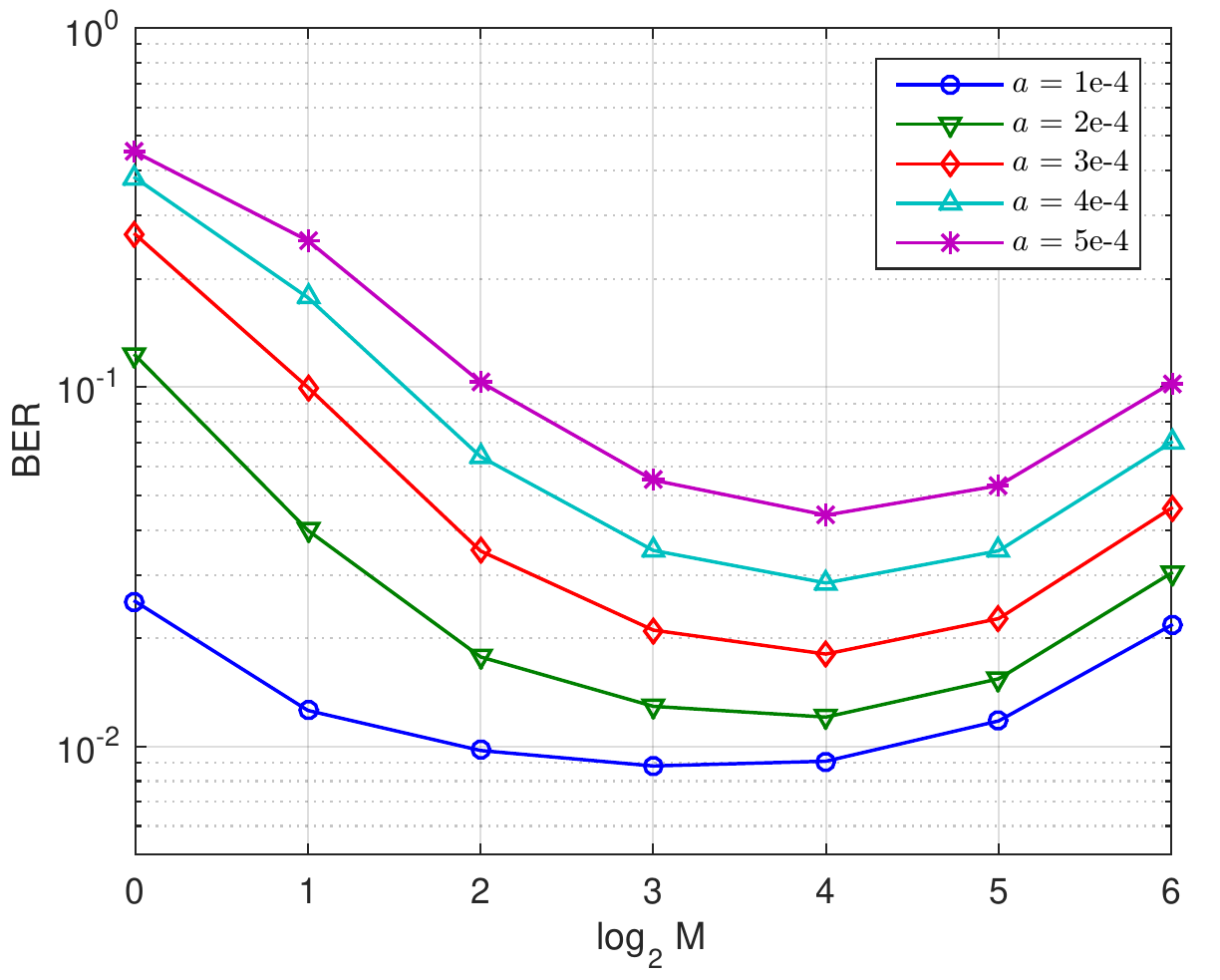}
\vskip -0.7em
\caption{BER performance of the proposed algorithm as a function of the number of subblocks $M$ with SNR fixed to $25$~dB.}
\label{fig3}
\vskip -1em
\end{figure}

Fig.~3 illustrates the BER performance of the proposed algorithm as a function of the number of subblocks $M$ for various values of the Doppler scaling factor $a$ ranging from $1 \times 10^{-4}$ to $5 \times 10^{-4}$. Here, the signal-to-noise ratio (SNR) is fixed at $25$~dB. And to guarantee $I \ge M$, the number of pilot symbols is set to $I = 128$.
It is interesting to observe that, similar to the partial-FFT demodulation methods for the coherent OFDM systems in~\cite{Yerramalli&Stojanovic&Mitra_IEEEJSP_2012, Jing&Lingling&Geert_IEEESPL_2016}, there exists performance saturation for the differential OFDM system as $M$ increases. An excess value of $M$ may lead to weight overfitting, i.e., tracking the differential noise in~(5) and the frequency response mismatch between adjacent subcarriers instead of the channel time variation, and thus impairs the system performance. As mentioned in Section~III, we usually choose $M = 4 \sim 32$ for practical use.

\begin{figure}[!t]
\centering
\includegraphics[width=3in]{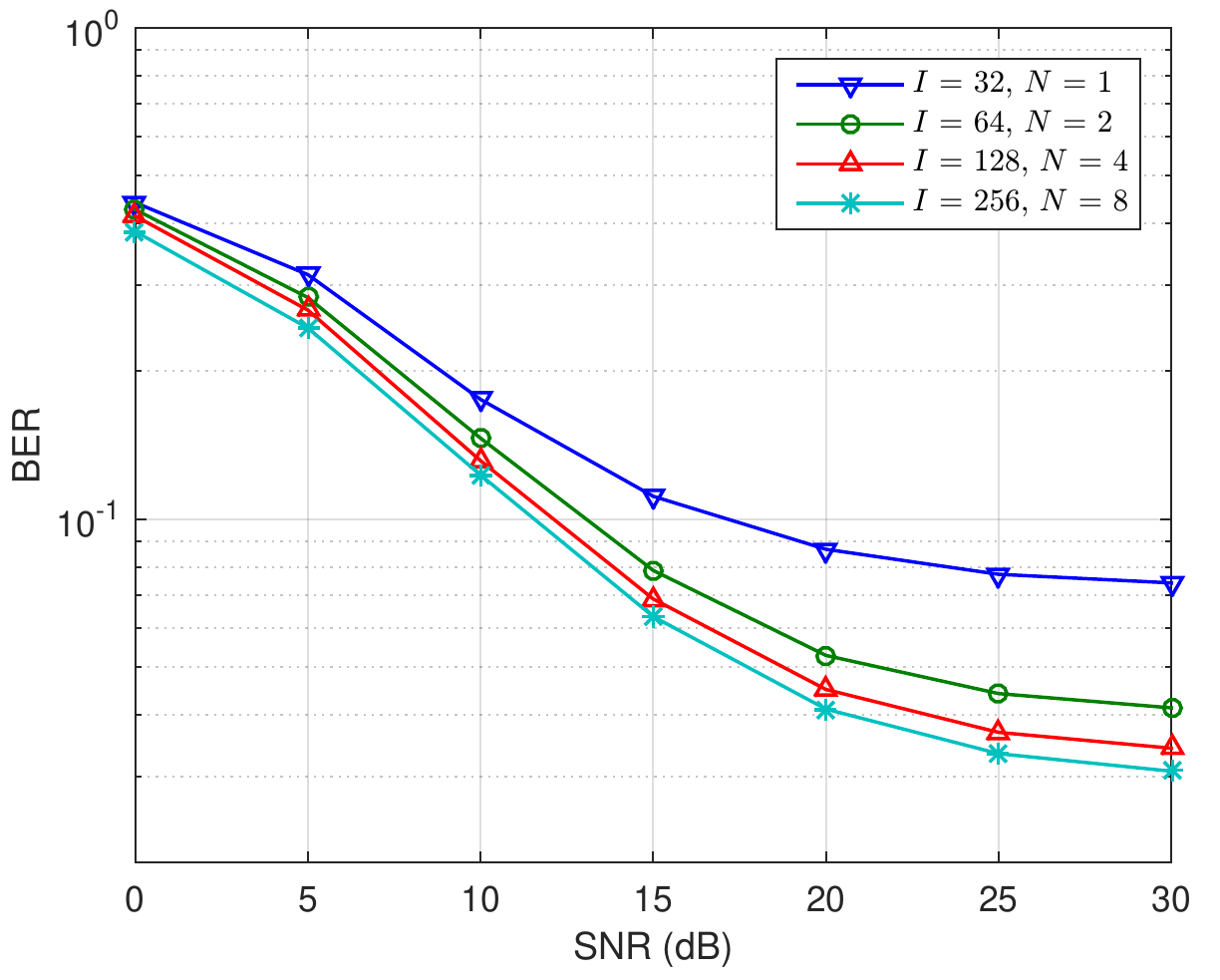}
\vskip -0.7em
\caption{BER performance of the proposed algorithm in the wideband scenario with the Doppler scaling factor $a = 5 \times 10^{-4}$.}
\label{fig4}
\vskip -1em
\end{figure}

Fig.~4 shows the performance of explicit wideband Doppler compensation of the proposed algorithm. As in Fig.~2, we fix the number of subblocks to $M = 8$. Also, to make the results more visible, the Doppler scaling factor is set to a relatively large value $a = 5 \times 10^{-4}$. In this case, the frequency shift at the lowest subcarrier of the differential OFDM system is about $\Delta f/ 2$, while about $\Delta f$ at the highest subcarrier; hence, the wideband Doppler effect is quite evident and should not be ignored.
To counteract its impact, we group the $K$ subcarriers into various numbers of subbands $N = 1,2,4$ and $8$, in each of which $\bar I = 32$ pilot symbols are allocated. As expected, the system performance improves as $N$ increases. However, it can also be observed that, the performance gain thus obtained becomes trivial when the number of subbands reaches a certain value ($N=4$ in this example). It implies that the narrowband assumption can now be held accurately enough in each subband. Therefore, from the practical point of view, no further increase in $N$ is needed, and a suitable value of $N$ can be easily determined in advance.

\section{Conclusions}

An eigendecomposition-based partial FFT demodulation algorithm is proposed in this paper for differential OFDM  to combat ICI over time-varying UWA channels.
The algorithm incurs only a moderate pilot overhead and low complexity.
Numerical simulation results demonstrate its performance superiority over existing adaptive methods and the rationale for choosing the algorithm parameters.

\appendices
%


\ifCLASSOPTIONcaptionsoff
  \newpage
\fi



\bibliographystyle{IEEEtran}
\bibliography{IEEEabrv,../bib/HJ}    

\begin{thebibliography}{10}
\providecommand{\url}[1]{#1}
\csname url@samestyle\endcsname
\providecommand{\newblock}{\relax}
\providecommand{\bibinfo}[2]{#2}
\providecommand{\BIBentrySTDinterwordspacing}{\spaceskip=0pt\relax}
\providecommand{\BIBentryALTinterwordstretchfactor}{4}
\providecommand{\BIBentryALTinterwordspacing}{\spaceskip=\fontdimen2\font plus
\BIBentryALTinterwordstretchfactor\fontdimen3\font minus
  \fontdimen4\font\relax}
\providecommand{\BIBforeignlanguage}[2]{{%
\expandafter\ifx\csname l@#1\endcsname\relax
\typeout{** WARNING: IEEEtran.bst: No hyphenation pattern has been}%
\typeout{** loaded for the language `#1'. Using the pattern for}%
\typeout{** the default language instead.}%
\else
\language=\csname l@#1\endcsname
\fi
#2}}
\providecommand{\BIBdecl}{\relax}
\BIBdecl

\bibitem{Lanbo&Shengli&Jun-Hong_WCommMCom_2008}
L.~Liu, S.~Zhou, and J.-H. Cui, ``Prospects and problems of wireless
  communication for underwater sensor networks,'' \emph{Wirel. Commun. Mob.
  Comput.}, vol.~8, no.~8, pp. 977--994, Aug. 2008.

\bibitem{Stojanovic_IEEEJOE_1996}
M.~Stojanovic, ``Recent advances in high-speed underwater acoustic
  communications,'' \emph{{IEEE} J. Ocean. Eng.}, vol.~21, no.~2, pp. 125--136,
  Apr. 1996.

\bibitem{Stojanovic&Preisig_IEEEMCOM_2009}
M.~Stojanovic and J.~Preisig, ``Underwater acoustic communication channels:
  Propagation models and statistical characterization,'' \emph{{IEEE} Commun.
  Mag.}, vol.~47, no.~1, pp. 84--89, Jan. 2009.

\bibitem{Yahong&Chengshan&TCYang_PHYCOMM_2010}
Y.~R. Zheng, C.~Xiao, T.~Yang, and W.-B. Yang, ``Frequency-domain channel
  estimation and equalization for shallow-water acoustic communications,''
  \emph{J. Phys. Commun.}, vol.~3, no.~1, pp. 48--63, Mar. 2010.

\bibitem{Zhendao&Giannakis_IEEEMSP_2000}
Z.~Wang and G.~B. Giannakis, ``Wireless multicarrier communications: Where
  {Fourier} meets {Shannon},'' \emph{{IEEE} Signal Process. Mag.}, vol.~17,
  no.~3, pp. 29--48, May 2000.

\bibitem{shengli&Zhaohui_OFDMbook_2014}
S.~Zhou and Z.~Wang, \emph{{OFDM} for Underwater Acoustic
  Communications}.\hskip 1em plus 0.5em minus 0.4em\relax John Wiley \& Sons,
  Jun. 2014.

\bibitem{Aval&Stojanovic_IEEEJOE_2015}
Y.~Aval and M.~Stojanovic, ``Differentially coherent multichannel detection of
  acoustic {OFDM} signals,'' \emph{{IEEE} J. Ocean. Eng.}, vol.~40, no.~2, pp.
  251--268, Apr. 2015.

\bibitem{Rugini&Banelli&Leus_IEEEJCOML_2005}
L.~Rugini, P.~Banelli, and G.~Leus, ``Simple equalization of time-varying
  channels for {OFDM},'' \emph{{IEEE} Commun. Lett.}, vol.~9, no.~7, pp.
  619--621, Jul. 2005.

\bibitem{KunFang&Rugini&Geert_IEEEJSP_2008}
K.~Fang, L.~Rugini, and G.~Leus, ``Low-complexity block turbo equalization for
  {OFDM} systems in time-varying channels,'' \emph{{IEEE} Trans. Signal
  Process.}, vol.~56, no.~11, pp. 5555--5566, Nov. 2008.

\bibitem{Schniter_IEEEJSP_2004}
P.~Schniter, ``Low-complexity equalization of {OFDM} in doubly selective
  channels,'' \emph{{IEEE} Trans. Signal Process.}, vol.~52, no.~4, pp.
  1002--1011, Apr. 2004.

\bibitem{Tu&Fertonani&Duman&Stojanovic_IEEEJOE_2011}
K.~Tu, D.~Fertonani, T.~M. Duman, M.~Stojanovic, J.~G. Proakis, and P.~Hursky,
  ``Mitigation of intercarrier interference for {OFDM} over time-varying
  underwater acoustic channels,'' \emph{{IEEE} J. Ocean. Eng.}, vol.~36, no.~2,
  pp. 156--171, Apr. 2011.

\bibitem{Jianzhong&Shengli&Jie_IEEEJSTSP_2011}
J.~Huang, S.~Zhou, J.~Huang, C.~Berger, and P.~Willett, ``Progressive
  inter-carrier interference equalization for {OFDM} transmission over
  time-varying underwater acoustic channels,'' \emph{{IEEE} J. Sel. Topics
  Signal Process.}, vol.~5, no.~8, pp. 1524--1536, Dec. 2011.

\bibitem{Baosheng&Shengli&Stojanovic_IEEEJOE_2008}
B.~Li, S.~Zhou, M.~Stojanovic, L.~Freitag, and P.~Willett, ``Multicarrier
  communication over underwater acoustic channels with nonuniform {Doppler}
  shifts,'' \emph{{IEEE} J. Ocean. Eng.}, vol.~33, no.~2, pp. 198--209, Apr.
  2008.

\bibitem{Kang&Iltis_IEEEJJSAC_2008}
T.~Kang and R.~A. Iltis, ``Iterative carrier frequency offset and channel
  estimation for underwater acoustic {OFDM} systems,'' \emph{{IEEE} J. Sel.
  Areas Commun.}, vol.~26, no.~9, pp. 1650--1661, Dec. 2008.

\bibitem{Yerramalli&Stojanovic&Mitra_SPAWC_2010}
S.~Yerramalli, M.~Stojanovic, and U.~Mitra, ``Partial {FFT} demodulation: A
  detection method for doppler distorted {OFDM} systems,'' Jun. 2010, pp. 1--5.

\bibitem{Yerramalli&Stojanovic&Mitra_IEEEJSP_2012}
------, ``Partial {FFT} demodulation: A detection method for highly {Doppler}
  distorted {OFDM} systems,'' \emph{{IEEE} Trans. Signal Process.}, vol.~60,
  no.~11, pp. 5906--5918, Nov. 2012.

\bibitem{Jing&Lingling&Geert_IEEESPL_2016}
J.~Han, L.~Zhang, and G.~Leus, ``Partial {FFT} demodulation for {MIMO-OFDM}
  over time-varying underwater acoustic channels,'' \emph{{IEEE} Signal
  Process. Lett.}, vol.~23, no.~2, pp. 282--286, Feb. 2016.

\bibitem{Stojanovic_SAM_2010}
M.~Stojanovic, ``A method for differentially coherent detection of {OFDM}
  signals on doppler-distorted channels,'' Oct. 2010, pp. 85--88.

\bibitem{Sharif&Neasham&Hinton_IEEEJOE_2000}
B.~S. Sharif, J.~Neasham, O.~R. Hinton, and A.~E. Adams, ``A computationally
  efficient {Doppler} compensation system for underwater acoustic
  communications,'' \emph{{IEEE} J. Ocean. Eng.}, vol.~25, no.~1, pp. 52--61,
  Jan. 2000.

\bibitem{Stojanovic&Catipovic&Proakis_IEEEJOE_1994}
M.~Stojanovic, J.~Catipovic, and J.~Proakis, ``Phase-coherent digital
  communications for underwater acoustic channels,'' \emph{{IEEE} J. Ocean.
  Eng.}, vol.~19, no.~1, pp. 100--111, Jan. 1994.

\bibitem{JHan&Leus_IEEEJOE_EA}
J.~Han, S.~P. Chepuri, Q.~Zhang, and G.~Leus, ``Iterative per-vector
  equalization for orthogonal signal-division multiplexing over time-varying
  underwater acoustic channels,'' \emph{{IEEE} J. Ocean. Eng.}, 2016, under
  review.

\bibitem{Matrix_Meyer_2000_inbook_MinMaxEigen}
C.~D. Meyer, \emph{Matrix Analysis and Applied Linear Algebra}.\hskip 1em plus
  0.5em minus 0.4em\relax Philadelphia, PA: SIAM, 2000, pp. 549--550.

\end{thebibliography}

\end{document}